# Optical-controlled wavelength-tunable Q-switched mode-locked fiber laser based on graphene-deposited micro-FBG

**Yujia Li, Lei Gao, Tao Zhu\*, Yulong Cao, and Jingdong Zhang**

*Key Laboratory of Optoelectronic Technology & Systems (Ministry of Education), Chongqing University, Chongqing 400044, China*
*Corresponding author: zhutao@cqu.edu.cn*

## Abstract

We report a wavelength-tunable Q-switched mode-locked fiber laser based on a compact optical tuning device, which is fabricated by coating the single-layer graphene on the surface of micro-fiber Bragg grating (MFBG). Owing to the evanescent wave interaction between graphene and MFBG, the center wavelength of MFBG can be accurately controlled by adjusting power of an external pump laser. By inserting the fabricated device into a compact fiber laser cavity mode-locked by single-wall carbon nanotubes, stable Q-switched mode-locked pulse is generated. Experimental results show that the wavelength can be tuned from 1550.1 nm to 1551.9 nm with a linear sensitivity of 6.3 pm/mW. Such an optical-controlled wavelength-tunable Q-switched mode-locked fiber laser may find potential applications in all-optical computing, all-optical logic gating and generation of microwave source.

## 1. Introduction

The study of wavelength-tunable Passively Q-switched mode-locked (QML) fiber laser has attracted much attention for its compact structure, low cost and easiness for output coupling [1–3]. Differing from the typical mode-locked laser, it outputs high repetition frequency pulse laser with intensity modulation, and the modulation frequency can be easily tuned by the change of pump power [4, 5]. Furthermore, compared with continuous mode-locked laser, the operation state of Q-switched laser is more robust and achieves self-start for easiness. Therefore, it has tremendous potential application values in optical signal process, optical fiber sensing and material processing [6, 7]. In recent years, with the rise of novel materials, more methods for passively Q-switched, mode-locked technology has been demonstrated, such as graphene, graphene oxide, and topological insulators [8–13]. However, the two-structure of those materials makes those saturable absorbers very sensitive to the light polarization and influences the stability of laser operation state. As a contract, single-wall carbon nanotube (SWCNT) is one kind of a one-dimension nonlinear optic material with the polarization-insensitive characteristics and wide operating bandwidth [14–18], and it is widely used as the saturable absorption component in the passively QML fiber laser for its low polarization-dependent loss [19].

Fiber Bragg gating (FBG) is widely utilized into optical communication, optical sensing and fiber laser, owing to its low cost, sensitive optical characteristics, pure fiber feature and steerable center wavelength and bandwidth [20]. To further enhance its sensitivity to the change of external physical field, the graphene-assited FBG has attracted many interests for the excellent optical characters of graphene, such as the optical Ker effect, saturable absorption and photo-thermal effect [21–24], and this device has been applied in the precise sensing for gas, temperature and stress [25–27]. In recent years, many control methods for bandwidth and center wavelength of FBG have been reported, including external mechanical stress, magnetic field, and thermal heating [28–30]. X. Tao et al. has demonstrated that the center wavelength of FBG could be all-optical controlled by injecting an additional modulation pump into the graphene-coated micro-FBG (GMFBG) [24]. Compared with the previous controlling methods, this low cost all-optical tuning device can achieve more accurate and faster wavelength tuning of FBG in a relatively small wavelength range for easiness by a compact tuning system. By the change of modulation pump power, an obvious red-shift of the center wavelength is observed, and it provides an efficient avenue for the all-optical tuning for the output wavelength of the QML laser. However, it suffers from the serious reflected spectral deterioration and strong broadening of bandwidth under the enhanced pump power, which may restrict its application for the wavelength tuning in laser system.

In our experiment, we have built the QML laser by combined the saturable absorber based on SWCNT-film and the all-optical tuning device fabricated via coating the single-layer graphene on the MFBG. The QML pulse laser train is generated inside a compact fiber laser cavity. To restrain the reflected spectral depravation, we coat the graphene film near the meddle region of MFBG with a length of 5.8 mm. When the pump power is changed, the repetition frequency of pulse train linearly varies from 6 KHz to 15 KHz with a change slope of 0.23 KHz/mW. By adjusting the modulation pump power, the precise all-optical wavelength tuning is accomplished with a range from 1550.1 nm to 1551.9 nm and a tuning sensitivity of 6.3 pm/mW. This wavelength-tunable QML fiber laser has tremendous potential value in all optical computing and generation of microwave source for its highly linear tuning performance and negligible wavelength return trip error.

## 2. Experimental principle and fabrication

The commercial single-mode uniform FBG used in our experiment has a reflectivity of 90% at 1550 nm and reflected bandwidth of 0.2 nm, and the length of its gating region is 1.2 mm. To enhance the evanescent wave escaping from the core, the MFBG is prepared by etching the FBG in the hydrofluoric acid with concentration of 8% for 6 hour and 40 minutes. To refrain from the large insertion of FBG, the selected diameter of FBG is 15.6 mm. Compared with the multi-layer graphene film, the single-layer graphene has weaker absorption characteristics and can enhance the uniformity of deposition. As shown in Fig. 1(a), the single-layer graphene is coated near the meddle region of gating region, and the length of deposition is 5.6 mm. Though the reflected spectral depravation can be better restrained with a longer coating region, meanwhile the stronger insertion loss makes this tuning device useless in the laser system. The insert of Fig. 1(a) shows spectra of FBG, MFBG and GMFBG. Due to the etched cladding, the original single mode structure of FBG is destroyed, the whole device can be regarded as a mix-wave guide consisted of single-mode and multi-mode structures. The attenuation must be generated resulted from the mode mismatch, when light passes the junction between the single-mode and multi-mode structure. The slight wavelength red-shift is mainly caused by the change of strain in the FBG during etched in hydrofluoric acid. When graphene is deposited on the MFBG, the wavelength blue-shift is also emerged. Compared with MFBG, the efficiency index of GMFBG is slightly enlarged [25]. However, the stress deformation of MFBG happened during the transfer process of graphene in deionized water may result in blue-shift of the center wavelength. The obvious variation of insertion loss is mainly caused by the serious absorption of graphene with such a long interaction length.

In order to study the influence on the mode-field distribution induced by Graphene, we calculate the LP01 mode of GMFBG at wavelengths of 1530 nm and 980 nm by the finite element method, respectively. As seen in Fig. 2, almost all the energy of LP01 mode is trapped inside the core of FBG. The ratio of the intensity in the center region and that on the junction between graphene and cladding is about 103, indicating a weak evanescent wave interaction between graphene and MFBG. This can ensure a good operation of this tuning device under a relatively high modulation pump power. The energy distribution at the wavelength of 1530 nm is more dispersive than that at 980 nm, together with stronger evanescent wave escaping from cladding, which may induce higher tuning sensitivity of FBG. The acute attenuation in the white part of inset indicates a strong absorption of graphene, especially for the wavelength of 1530 nm. Thus, we have to carefully optimize the length of graphene-deposited region of MFBG to decrease the insertion loss of GMFBG.

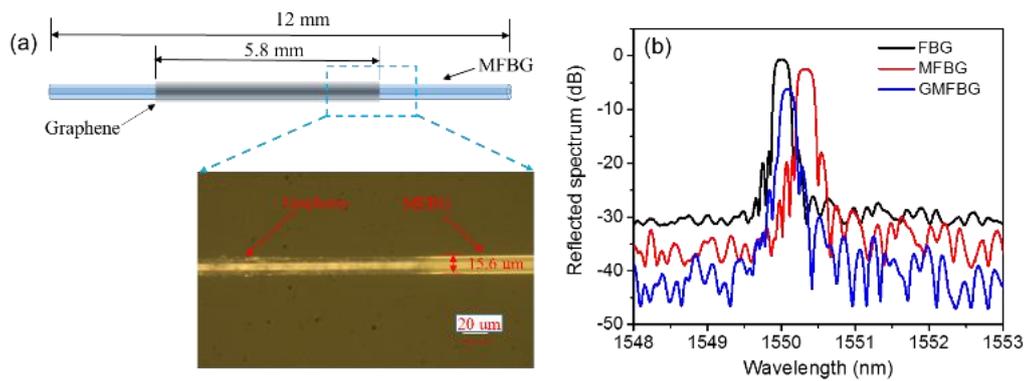

**Fig. 1.** (a) The diagram schematic of GMFBG, and the inset is the photograph of junction region of GMFB and MFBG. (b) The reflected spectra of FBG, MFBG and GMFBG.

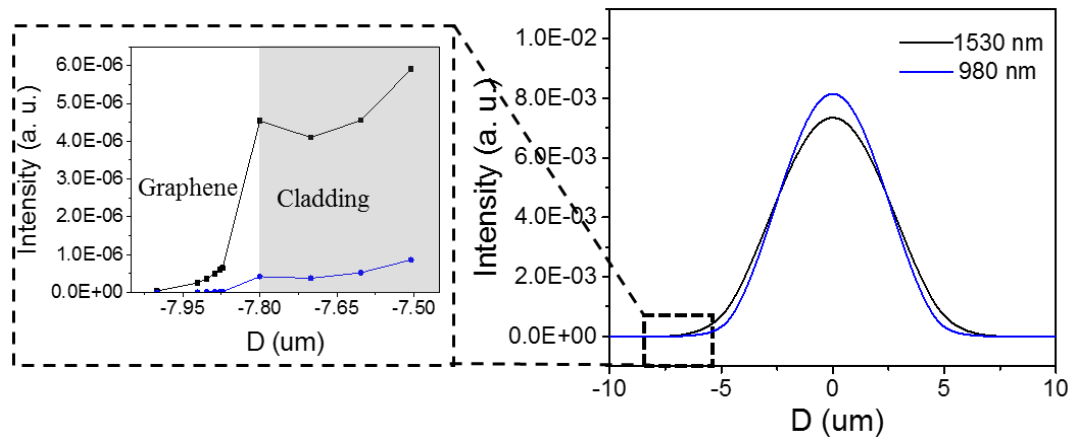

**Fig. 2.** The one-dimension mode distribution of a cutting line passing through the center of GMFBG at wavelengths of 1553 nm and 980 nm, and the inset inside the black dashed box is the partial mode-field distribution near by the junction of graphene and cladding. The white and gray areas represent graphene and cladding of MFBG, respectively.

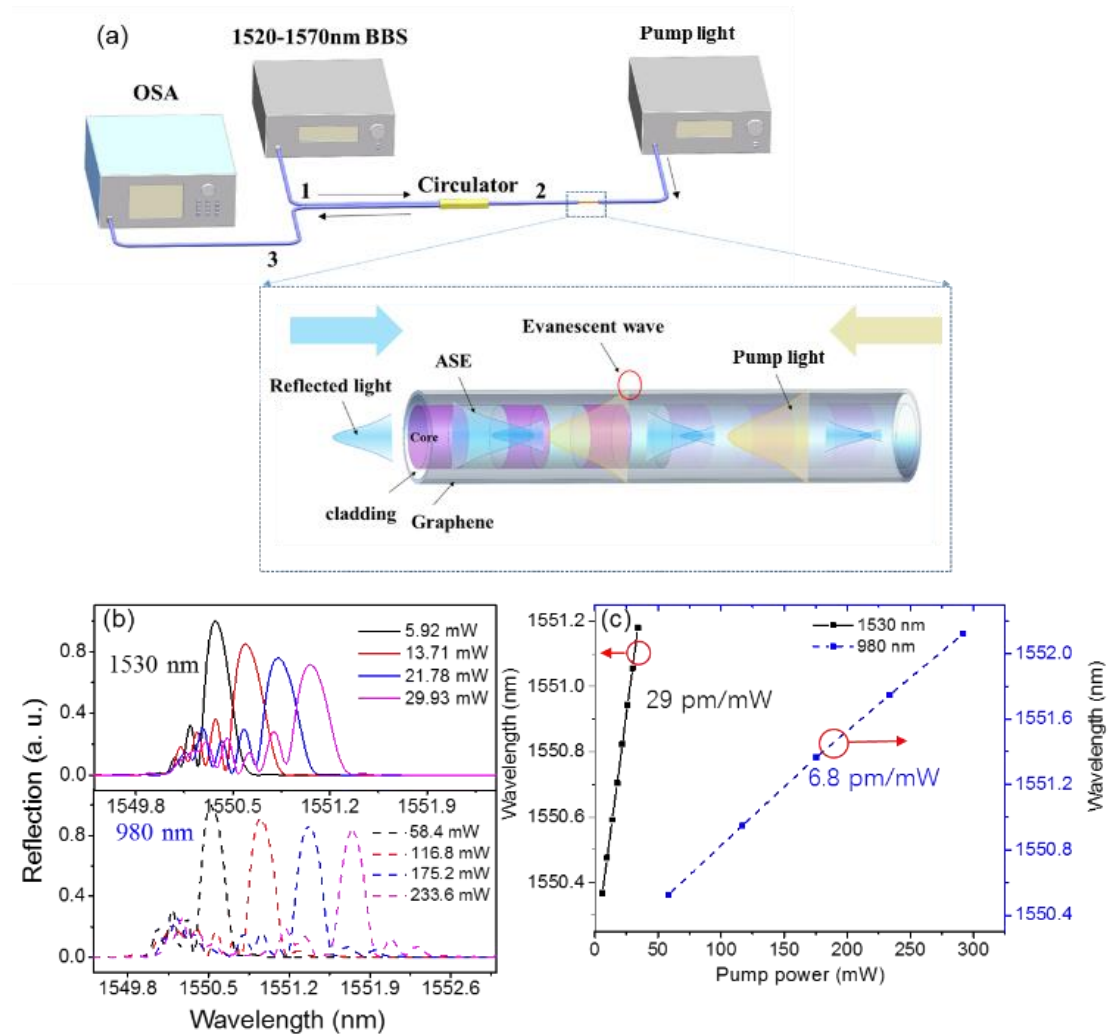

**Fig. 3.** (a) The test system of the GMFBG, and the inset inside the blue dashed box shows the process of evanescent wave interaction between graphene and MFBG. The yellow and blue arrows represent the directions of pump and signal light that enter into the MFBG, respectively. (b) The reflected spectra under different pump powers at wavelengths of 1530 nm and 980 nm. (c) Reflected peak wavelengths under different pump powers at the corresponding wavelengths, and the tuning sensitivity of 1530nm and 980 nm are 29 pm/mW and 6.8 pm/mW.

To demonstrate the optical-controlled performance of the fabricated GMFBG, we build a test system for the wavelength tuning as shown in Fig. 3(a). We use a broad bandwidth source (BBS) with a wavelength range from 1520 nm to 1570 nm as the signal light, which is injected into the 1 port of a circulator. Because graphene has a wide operation bandwidth, this tuning device need not a special selection for the wavelength of pump. The 980nm and 1530 nm pump lasers utilized as the modulation pump light are directly injected into the other port of GMFBG, respectively. The reflected spectrum of GMFBG is observed by an optical spectrum analyzer (OSA). The inset inside the blue dashed box shows the interaction between evanescent wave and graphene. The pump laser, which is far from the reflected bandwidth of FBG, will directly pass through GMFBG. Figure 3(b) shows the reflected spectra of GMFBG under different modulation pump powers at wavelengths of 1530 nm and 980 nm. The obvious red-shift of wavelength is obtained by increasing the pump power. However, the reflectivity of FBG is decreased, together with the spectrum deteriorated. The side-lobes at left edge of the main-lobes are generated. The uneven distribution of graphene deposited on the MFBG will influence the periodical index structure of mix-wave guide. Some wavelengths far from the main reflected peak may also be reflected. With the increase of pump power, both the side-lobes and main-lobe will have red shift. Even though those wavelengths cannot satisfy the reflected phase match condition of original period waveguide of FBG. GMFBG under pump light at the wavelength of 1530 nm has a more serious spectral deterioration and insertion loss, due to the more decentralized mode distribution and stronger absorption induced by graphene, coinciding with the results of simulation. When the pump light interacts with graphene, the distribution of carriers in the energy band can be changed for easiness through absorbing pump photons, resulted from the zero-gap structure of graphene. Considering the high thermal conductivity of graphene, the random transition of carriers, which is induced by the photo-thermal effect, cannot be ignored under a pump power of several hundred milliwatts, and it will have a considerable influence on the index of graphene. We have known that the reflected center wavelength of FBG is determined by the effective index. When we increase the pump power, this change of carrier distribution will directly change the effective index of the mix-wave guide, giving rise to considerable red-shift of the reflected center wavelength. Thus, the wavelength-shift can be simply expressed by the following formula:

where $\Lambda$ is the index modulation period, and P and neff are defined as the power of modulation pump and effective index of this fabracated mix-wave guide, respectively. Owing to the high thermal threshold of GMFBG with evanescent wave interaction, this device even can tolerate a pump power of nearly 300 mW. Figure 3(c) depicts the linear relationship between the reflected peak wavelength and pump power at wavelengths of 1530 nm and 980 nm. Owing to a stronger evanescent wave interaction with pump wavelength of 1530 nm, more light contacts with graphene, resulting in a higher sensitivity. Therefore, the above discussion indicates that the intensity of evanescent wave is a critical factor to determine the tuning sensitivity. However, higher sensitivity will also induce worse spectral deterioration.

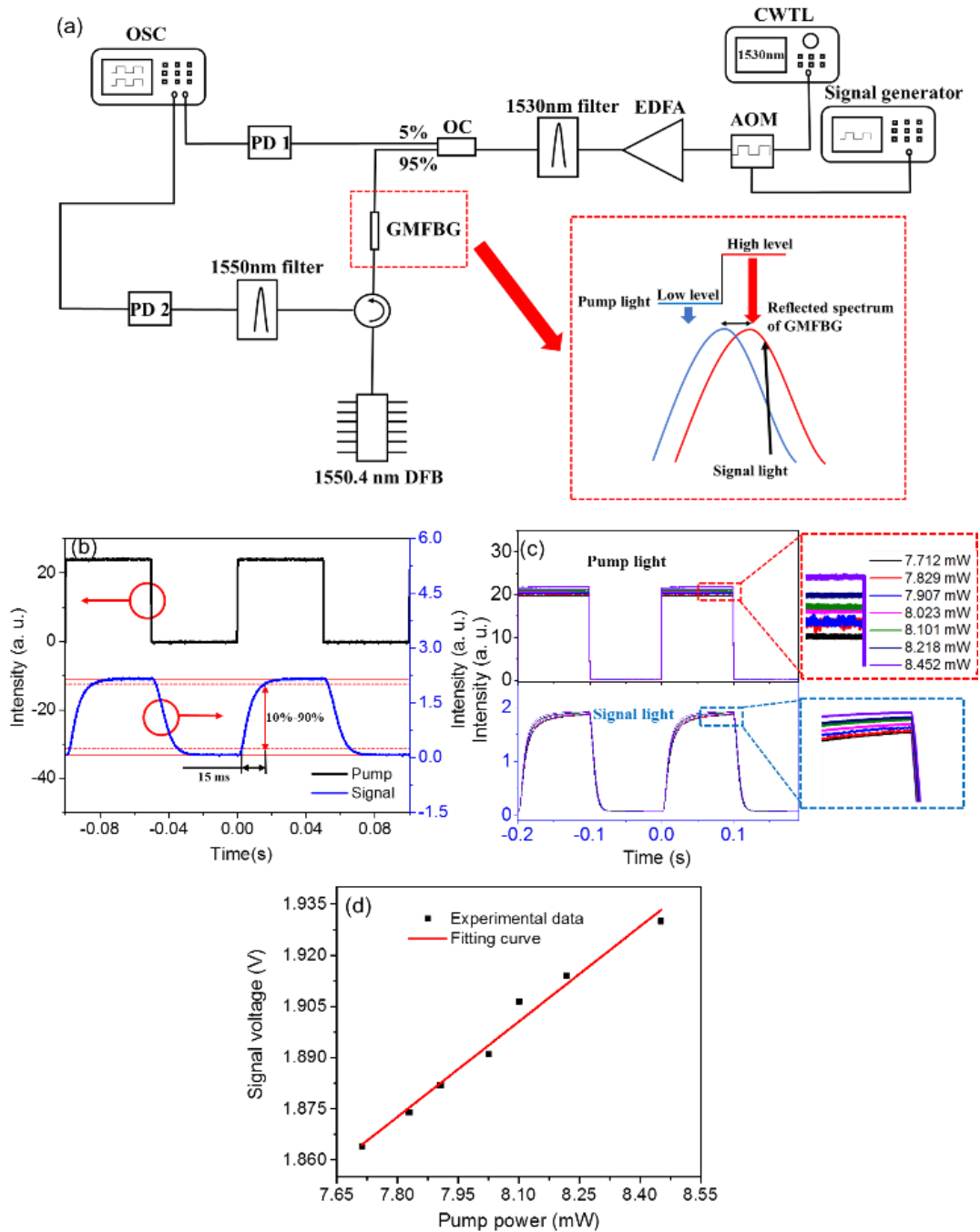

**Fig. 4.** (a) The test system for response time and tuning precision of GMFBG. (b) The temporal electric signals of pump light and signal light with a modulation of 10 Hz and duty circle of 50%. (c) The temporal electric signals of pump light and signal light with pump power slightly changed. (d)The highest level of signal voltage versus different pump powers.

As seen in Fig. 4(a), the pump laser is modulated to be square wave pulse with a duty circle of 50%. A 1550.4 nm narrow linewidth laser is utilized to be the signal light. As shown in the inset of red dashed box, the red and blue curves represent the reflected spectrum under pump light with high and low levels, respectively. In order to enhance the modulation depth, the signal light is fixed at the falling edge of the reflected spectrum shown as the inset with the red dashed box. Therefore, the bandwidth of GMFBG is periodically switched, which causes an intensity modulation for the direct-current signal light. Figure 4(b) depicts the isochronous measurement of modulated pump light and synchronous signal light. The response time is calculated for signal changed from 10% to 90% of high level shown as the red double arrow. Thus, the response time of the tuning device is about 15 ms. It is very difficult to directly measure the minimum tuning step size by an OSA for the limited wavelength resolution. As shown in Fig. 4(c), we just slightly enhance the power of pump light with a fixed modulation frequency, and the relative position of signal light with respect to the reflected spectrum will be changed, leading to an increment of highest level of signal light. The inset with the red dashed box is an enlargement of pump light with the power from 7.712 mW to 8.452 mW. The inset with the blue dashed box shows the corresponding voltage of signal light near the highest level. Fig. 4(d) shows the highest level of signal light with the increment of the pump power. The average step size of pump light is about 0.123 mW. Thus, the tuning precision is experimentally eliminated by the following expression:

## 3. Experimental setup

As seen in Fig. 5, the outstanding optical-controlled wavelength-tunable performance of GMFBG is utilized to build a wavelength-tunable passively QML fiber laser. A 980 nm continuous pump laser is inserted into the laser cavity by a 980/1550 wavelength division multiplexer (WDM). We use 1m erbium-doped fiber (EDF, E08-A352A-01-1B22) with a dispersion coefficient of 15 ps/nm/km as the gain medium. A polarization controller is located behind the EDF to adjust the polarization-dependent loss and stochastic dispersion of the cavity. A polarization-independent isolator is used to determine the operation direction of laser system and protect the pump laser against the back reflected light induced by other devices of the cavity. A 90/10 optical coupler (OC) is inserted into this cavity to achieve the output and feedback of laser. The fabricated SWCNT-SA connected with the OC is a key component to periodically modulate the net cavity loss and generate the Q-switched envelope of mode-locked pulse. SWCNT-film is prepared by mixing 10 wt% aqueous polyvinyl alcohol (PVA) and 0.5 mg/mL SWCNTs at a volume ratio of 1:2 for several hours by a high power blender. Then the mixture is placed in a thermostatic drying box until it becomes a semitransparent film. SWCNT-SA is fabricated by sandwiching the SWCNT-film between two fiber connectors. As seen from Fig. 3(b), compared with pump light at a wavelength of 1530nm, 980 nm pump has a milder reflected spectral deterioration and lower insertion loss within the same tuning range. Besides, we have known that the circulator with operation bandwidth at C band would filter the 980 nm pump laser. Thus, the modulation pump laser will only influence the tuning device and not enter the laser cavity to cause crosstalk to the signal laser. Therefore, another 980 pump laser is used as the modulation pump for the wavelength tuning of the reflected band of GMFBG. We use a circulator to connect the pump laser and GMFBG with the ring laser cavity.

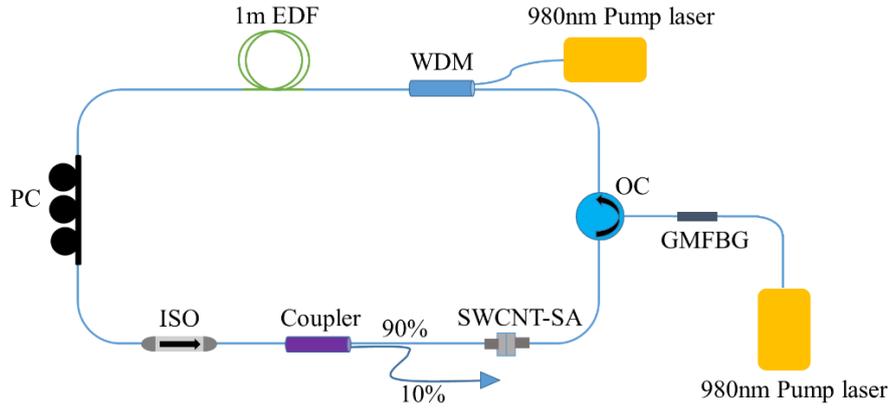

**Fig. 5.** The experimental schematic of GMFBG based wavelength-tunable passively QML fiber laser.

As seen in Fig. 4, we utilize the outstanding optical-controlled wavelength-tunable performance of GMFBG to build a wavelength-tunable passively QML fiber laser. A 980 nm continuous pump laser is inserted into the laser cavity by a 980/1550 wavelength division multiplexer (WDM). We use 1m erbium-doped fiber (EDF, E08-A352A-01-1B22) with a dispersion coefficient of 15 ps/nm/km as the gain medium. A polarization controller is located behind the EDF to adjust the polarization-dependent loss and stochastic dispersion of the cavity. A polarization-independent isolator is used to determine the operation direction of laser system and protect the pump laser against the back reflected light induced by other devices of the cavity. A 90/10 optical coupler (OC) is inserted into this cavity to achieve the output and feedback of laser. The fabricated SWCNT-SA connected with the OC is a key component to periodically modulate the net cavity loss and generate the Q-switched envelope of mode-locked pulse. SWCNT-film is prepared by mixing 10 wt% aqueous polyvinyl achohol (PVA) and 0.5 mg/mL SWCNT at a volume ratio of 1:2 for several hours by a high power machine. Then the mixture is placed in a thermostatic drying box until it becomes semitransparent film. SWCNT-SA is fabricated by sandwiching the SWCNT-film between two fiber connectors. Another 980 nm continuous wave laser is used as the modulation pump for the wavelength tuning of the reflected band of GMFBG. We use a circulator to connect the pump laser and GMFBG with the ring laser cavity. We have known that the circulator with operation bandwidth at C band would filter the 980 nm pump laser. Thus, the modulation pump laser will only influence the tuning device and not enter the laser cavity to cause crosstalk to the signal laser.

### 4. Experimental results and discussion

When we increase the pump power from 0 mW to 120 mW, the corresponding average output power is shown as Fig. 6(a). By properly adjusting the PC to optimize the total cavity loss and polarization-dependent loss, the stable QML laser pulse can be obtained, resulted from the excellent saturable absorption effect of SWCNT-SA and large loss of GMFBG induced by the narrow reflected bandwidth. The QML threshold is about 71 mW, due to the narrow filter performance of FBG. When the pump power is beyond the laser threshold, the output power is linearly increased with an enhanced pump power. Figure6(b) shows the spectrum of output laser with a center wavelength at 1550.1040 nm with the pump power of 99 mW, and the 3 dB bandwidth is merely 0.02 nm, which is much narrower than the reflected bandwidth. The QML laser is generated by adding an intensity modulation to the typical mode-locked pulse. Therefore, the stable QML pulse train includes both high frequency components corresponding to the mode-locked pulse and low frequency ingredients corresponding to the Q-switched envelope.

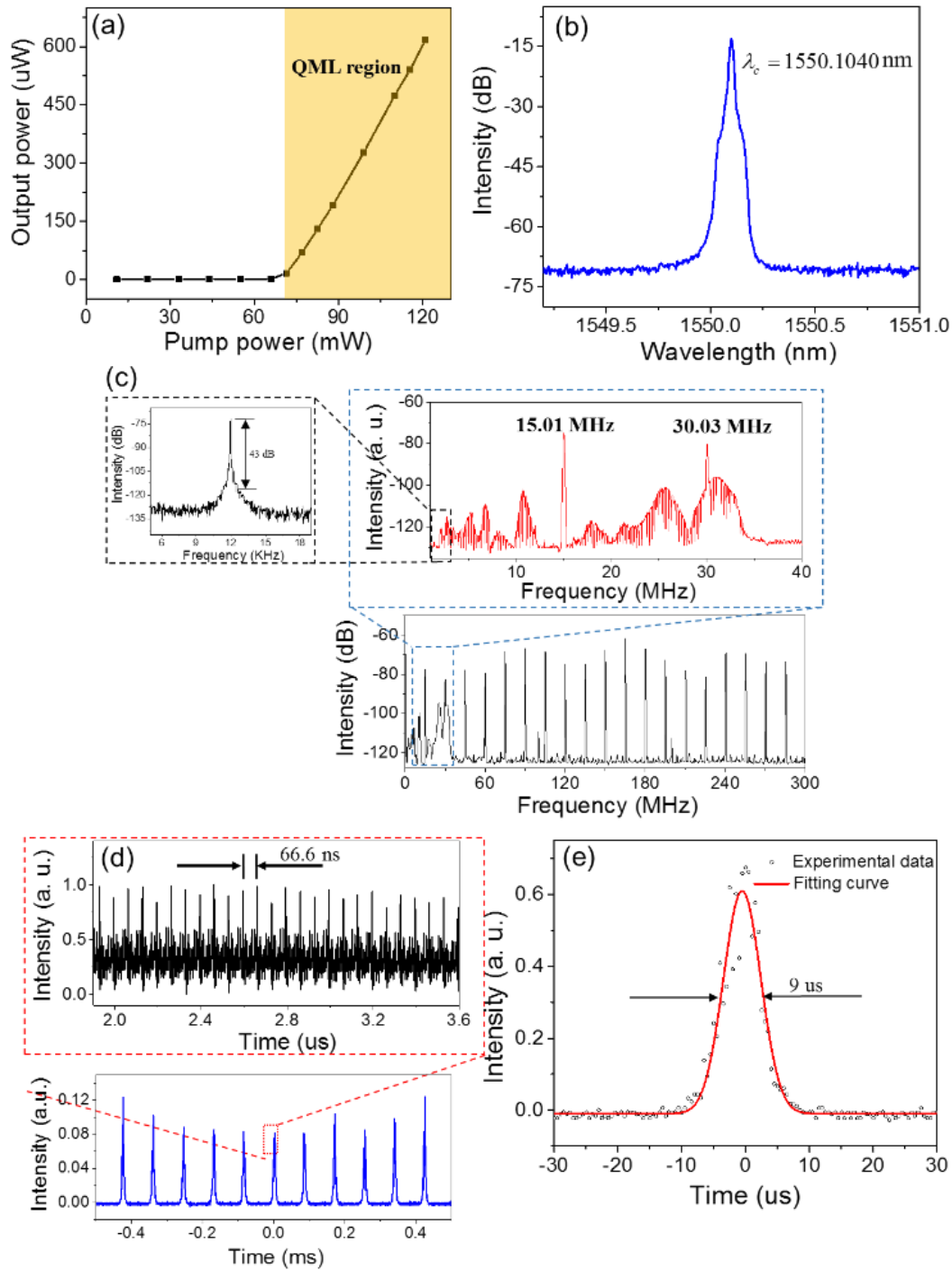

**Fig. 6.** (a) The output power under different pump powers. (b) The spectrum of output laser with a pump power of 99 mW. (c) The RF spectrum with a frequency span of 300 MHz, and the inset in the blue dashed box is the RF spectrum with a span of 40 MHz. The inset within the black dashed box shows the fundamental component at 12.01 KHz. (d) The Q-switched envelope pulse train in time domain, and the inset in the red dashed box is the mode-locked pulse train. (e) The duration of Q-switched envelope.

Figure 6(c) is the radio frequency (RF) spectrum of output laser with a span range of 300 MHz and resolution bandwidth of 50 Hz. The frequency space of 15.01 MHz is identical to the repetition frequency of fundamental mode-locked pulse, and this value is only decided by the total length of the laser cavity. However, as seen from the insert with the blue dashed box, besides high frequency components, there are also many frequency ingredients with smaller interval, which corresponds the low frequency Q-switched envelope. The inset within the back dashed box shows the fundamental frequency component at 12.09 KHz with a sweep span of 20 KHz and a resolution bandwidth of 5 Hz, where the signal to noise ratio of this component is about 43 dB, indicating a stable operation state of the QML laser. The corresponding pulse train in time domain is shown as Fig. 6(d), where the obvious Q-switched envelope is obtained with a period of 82.7 us, which coincides to the fundamental component. In order to further observe the mode-locked pulse, we measure the pulse train with a smaller time range. As seen in the inset of Fig. 6(d) inside the red dashed box, the period of mode-locked is 66.6 ns, corresponding to the frequency space of high frequency components in the Fig. 6(c). The duration of Q-switched envelope directly measured by the oscilloscope is about 9 us, which can be controlled by adjusting the pump power for easiness, and it is mainly determined by the switch time of the cavity Q-factor. With a higher pump power, the envelope duration can be further suppressed for the shorter switch time.

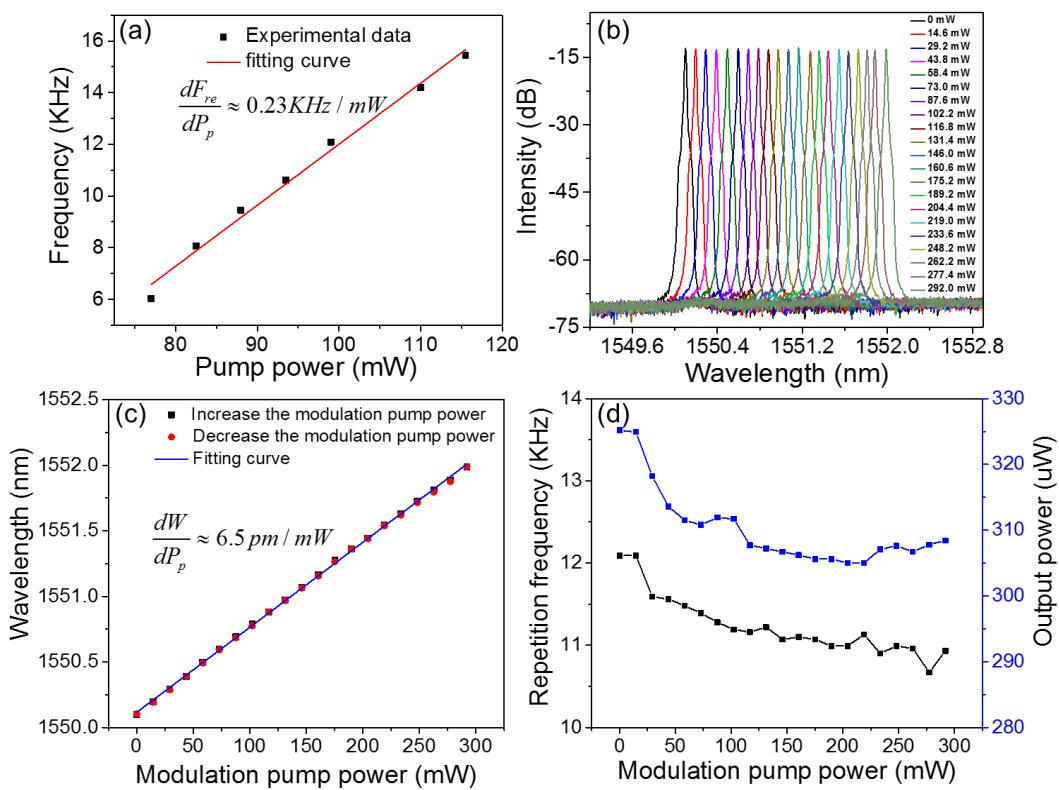

**Fig. 7.** (a) The repetition frequency of Q-switched envelope with different pump powers, and the red curve is the linear fitting curve. (b) The spectra of output laser with pump power with range from 0 mW to 292 mW. (c) The peak wavelengths of output laser at different pump powers with two change process of modulation pump power, and the experimental data is linearly fitted. (d) The repetition frequency and output power with different modulation pump powers.

Differing from the period of mode-locked pulse determined by the total length of laser cavity, the repetition frequency of Q-switched pulse is linearly increased by adjusting the pump power from

77 mW to 115.5 mW with a tuning slope of 0.23 KHz/mW, as seen in Fig. 7(a). With the enhancement of pump power, the pumping efficiency for gain fiber is improved, causing that the time for accumulation of inversion population is shortened, and the number of output pulse per unit time is increased. We keep the pump power of 180 mW unchanged and just adjust the modulation pump power from 0 mW to 292 mW to achieve the optical wavelength-tuning. As seen in Fig. 7(b), the wavelength tuning range from 1550.104 nm to 1552.1 nm is obtained without spectral distortion and any obvious minor lobes. In spites of the enhancement of side-lobes of reflected spectrum with a relatively high pump power, the height of side-lobes do not exceed one third of the main reflected. The distortion of reflected spectrum will not influence the wavelength tuning performance of this fabricated device, owing to the intrinsic mode competition of Er-dope fiber. Figure 7(c) shows the peak wavelengths under different modulation pump powers. The back splashes and red splashes represent the process of continuous increment and decrement of the pump power, respectively, and the return trip variance between those two progresses approaches to the spectral resolution limitation of 0.04 nm. Another reason is mainly due to the exited power jitter of pump laser. Therefore, by simply adjusting the power of modulation pump, the excellent optical-controlled linear tuning performance of pulse laser is obtained, and the wavelength-tuning sensitivity is 6.5 pm/mW with a linear fitting R2 of 0.99954, indicting a high wavelength-tuning precision. As the linear tendency of wavelength red-shift is unchanged during the whole tuning process, together with no thermal damage of GMFBG appearing, the tuning range can be further improved by enhancing the maximum power of modulation Pump. As seen in Fig. 7(d), both the repetition frequency and output power at different peak wavelengths are just moderately decreased accompanied with the redshift of the peak wavelength, where the variation range of repetition frequency and output power are 1.19 KHz and 20 uW, respectively. We have known that the pumping efficiency could influence the output power and repetition frequency of Q-switched envelope to a large extent. Due to the reducing of reflected peak of GMFBG shown in Fig. 3(b), the insertion loss of this tuning device is slight enhanced, which results in the decrement of pumping efficiency for the whole laser systems.

## 5. conclusion

We have proposed an optical wavelength-tunable passively QML fiber laser based on a MFBG deposited by graphene. With the excellent saturable absorption characteristics of SWCNTs, the stable QML laser pulse is oscillated inside a compact all-fiber laser cavity. By properly changing the power of the modulation pump, the peak wavelength of laser has an accurate red-shift from 1550.1040 nm to 1552.1 nm with a tuning slope of 6.3 pm/mW, together with the negligible return trip error. The linear fitting R2 of wavelength tuning can reach 0.99957. Due to the unchanged linear change tuning tendency during the whole tuning process and high thermal threshold of GMFBG, the variation range of wavelength can be further improved with a larger modulation pump power. This fabricated all-optical tuning device is firstly utilized in a fiber laser system to achieve optical-control tuning of laser wavelength, which may provide tremendous potential applications in all-optical computing, precise optical bioimaging and generation of microwave signal.